\documentclass[11pt, sf]{iopart}
\usepackage[english]{babel}
\usepackage[T1]{fontenc}
\usepackage{amssymb,amstext,amsbsy}
\usepackage{graphicx}
\usepackage{geometry}
\usepackage{url} 
\usepackage{texdraw}
\usepackage{xspace}
\usepackage{setspace}
\usepackage{rotate} 
\usepackage{psfrag}
\usepackage[labelfont={sf,bf,footnotesize}, textfont={footnotesize}]{caption}
\usepackage{placeins}

\usepackage{color}
	 \definecolor{darkred}{rgb}{0.75,0,0}
	 \definecolor{darkgreen}{rgb}{0,0.5,0}
	 \definecolor{darkblue}{rgb}{0,0,0.75}

\newcount\hour \newcount\minute
\hour=\time \divide \hour by 60 \minute=\time \count99=\hour
\multiply \count99 by -60 \advance\minute by \count99 \hour=\time
\divide \hour by 60

\date{\today}

\begin{document}

\title{Nonequilibrium phase transitions in finite arrays of \\
globally coupled Stratonovich models: Strong coupling limit
}
\author{Fabian Senf$^{[1,2]}$, Philipp M.~Altrock$^{[1,3]\ast}$, and Ulrich Behn$^{[1]}$}
\address{
$[1]$ Institute for Theoretical Physics, University of Leipzig,\\ \hspace{0.5cm}POB 100 920, D-04009 Leipzig, Germany\\
$[2]$ Present address: Leibniz-Institute of Atmospheric Physics,\\ \hspace{0.5cm}Schlo\ss stra\ss e 6, D--18225 K\"uhlungsborn, Germany\\
$[3]$ Present address: Max-Planck-Institute for Evolutionary Biology,\\ \hspace{0.5cm}August-Thienemann-Str. 2, D--24306 Pl\"on, Germany\\
Email: senf@iap-kborn.de, altrock@evolbio.mpg.de, and behn@itp.uni-leipzig.de\\
$\ast$ Author to whom correspondence should be addressed.
}
\date{\today}

\begin{abstract}
A finite array of $N$ globally coupled Stratonovich models exhibits a
continuous nonequilibrium phase transition. In the limit of strong coupling
there is a clear separation of time scales of center of mass and relative
coordinates. The latter relax very fast to zero and the array behaves as a
single entity described by the center of mass coordinate. We compute
analytically the stationary probability and the moments of the center of mass
coordinate. The scaling behaviour of the moments near the critical value of
the control parameter $a_c(N)$ is determined. We identify a  crossover from
linear to square root scaling with increasing distance from $a_c$. The
crossover point approaches $a_c$ in the limit $N \to \infty$ which reproduces
previous results for infinite arrays. The results are obtained in both the
Fokker-Planck and the Langevin approach and are  corroborated by numerical
simulations. For a general class of models  we show that the transition
manifold in the parameter space depends on $N$ and is determined by the
scaling behaviour near a fixed point of  the stochastic flow.
\end{abstract}
\maketitle

\section{Introduction}

Arrays of stochastically driven nonlinear dynamical systems may exhibit
nonequilibrium phase transitions of continuous or discontinuous type, for a
recent review see \cite{SSG-O07}, cf. also \cite{Munoz2004,GS99}. Concepts
developed to describe equilibrium phase transitions  such as
symmetry or ergodicity breaking, order parameter, critical behaviour, critical
exponents etc. have been successfully transfered to noise induced
nonequilibrium phase transitions. The structure of the theory will be 
generically of  mean field type, if infinite globally coupled arrays are
studied which allows for a number of analytical results.

Remarkably, essential characteristics of phase transitions can already be
found in the case of a  single Stratonovich model. This is mainly due to the
multiplicative nature of the noise. Models driven by additive noise do not
show this peculiar property.  The Langevin equation for the single-site
Stratonovich model
\cite{HL84,SB79,SMKG82} reads
\begin{equation}\label{1LE}
dx=(ax-x^3)dt +\sigma x \circ d W(t),
\end{equation}
where $a$ is a control parameter, $\sigma$ denotes the strength of the noise
and $W(t)$ is a Wiener process with autocorrelation $\langle W(t) W(s)
\rangle = \min{(t,s)}$. Equation (\ref{1LE}) is interpreted in
the Stratonovich sense as indicated by the symbol $\circ$. The Stratonovich
model describes, e.g., the overdamped motion in a biquadratic potential $U(x)= -
\frac {a}{2} x^2 + \frac {1}{4}x^4$ where the control parameter is
stochastically modulated, $a \to a +\xi_t$,  with a Gaussian white
noise $\xi_t$. 

The associated Fokker-Planck equation (FPE) describing the evolution of the
probability density $P(x,t)$ is
\begin{eqnarray}\label{1FPE}
\partial_t P = - \partial_{x} \Big\{\big[(a-\frac{\sigma^2}{2})x - x^3- 
\frac{\sigma^2}{2} x^2 \partial_{x}\big]P\Big\}.
\end{eqnarray}
Equation (\ref{1FPE}) has a weak stationary solution, a Dirac distribution
$\delta(x)$ located at the common zero $x=0$ of drift and diffusion
coefficient, which is also a zero of the stochastic flow in Eq. (\ref{1LE}).
If the system is initially at $x=0$ it will always stay there. 

Furthermore, there exist spatially extended strong stationary solutions
determined up to a constant factor, $P_s(x) \propto |x|^{2a/\sigma^2 -1}
\exp\{-(x/\sigma)^2\}$. $P_s(x)$ will live on $S_{+}=[0,\infty)$ if the initial
distribution lives on $S_+\backslash 0$, and on $ S_-=(-\infty,0]$, if the
initial distribution lives on $S_-\backslash 0$. The constant is determined
such that the solution is normalized integrating over the support and can
be interpreted as probability density, i.e.
\begin{eqnarray}
P_s(x)=\frac{1}{Z}|x|^{2a/\sigma^2 -1} e^{-(x/\sigma)^2},\label{SM1P}\\
Z =\int_{S_{\pm}} dx |x|^{2a/\sigma^2 -1}e^{-(x/\sigma)^2}=
\frac{1}{2}\sigma^{2a/\sigma^2}\Gamma (a/\sigma^2), \label{SM1Z}
\end{eqnarray}
provided $ 2a/\sigma^2 >0$. For $ 2a/\sigma^2 \leq 0$ the normalization $Z$
diverges since the integrand in (\ref{SM1Z}) scales like $|x|^{2a/\sigma^2
-1}$ as $x \to 0.$ In this case it can be shown (see \ref{StratIto}) that a weakly
normalized version converges to the known weak solution $\delta(x)$ and $x=0$ is
an absorbing fixed point of the system.

If fractions of the initial distribution  of given weights live on $S_-$, on
$S_+$, and on $0$, all will keep their weight and evolve to the stationary
probability densities living on their respective support as guaranteed by a
$H$-theorem \cite{SB_Htheorem}.

The Stratonovich model exhibits  a strong ergodicity breaking \cite{BB06}
depending on the control parameter $a$, since the state space decomposes into
regions where the system cannot reach one region if it has started in a
different one. For $a\leq 0$ the only stationary solution is $\delta(x)$, i.e.
the fixed point $x=0$ of the stochastic dynamics is absorbing. Additionally,
for $a>0$ we have the spatially extended solution (\ref{SM1P}) living on
$S_\pm$ depending on the initial distribution. This is reflected by the mean
value
\begin{equation}
\langle x \rangle_\pm = \int_{S_\pm} \!\!dx x P_s(x) = 
\cases
{
0 \hspace{3.5cm}\text{if} \;a \leq 0,\\
\pm \sigma
\frac{\Gamma({a}/{\sigma^2}+{1}/{2})}{\Gamma({a}/{\sigma^2})} \hspace{0.5cm}\text{if} 
\;a > 0 .
}
\end{equation}
Obviously, $\langle x \rangle_\pm$ can serve as an order parameter and shows a
critical behavior $\langle x \rangle_\pm \sim \pm \frac
{\sqrt{\pi}}{\sigma}(a-a_c(1))^\beta$ as $a \to a_c(1)=0$ with  $\beta =1$. 

Note that also the location of the maximum of the spatially extended density
undergoes a bifurcation, $x^{\text{max}}_\pm =0$ for $0 < a \leq
a_c^{\text{max}}(1)=\sigma^2/2 $ and $x^{\text{max}}_\pm = \pm
(a-a_c^{\text{max}})^{1/2}$ for $a \geq a_c^{\text{max}}(1)$.

The critical behaviour of an array of {\it infinitely} many globally coupled
Stratonovich models has been thoroughly investigated in \cite{BLMKB02}. The
scaling of higher moments was considered in \cite{MCC05}, see also
\cite{Przybilla02}. The stationary probability density is the solution of a
nonlinear Fokker-Planck equation which depends on the order parameter.  The
scaling behaviour of the order parameter  is analytically  determined,
$\langle x \rangle_\pm \sim \pm (a-a_c(\infty))^\beta$ as $a \to a_c(\infty)$
with $a_c(\infty)= -\sigma^2/2$ and $\beta = \sup \{1/2, \sigma^2/(2D)\}$,
where $D$ is the strength of the harmonic coupling between the systems
\cite{BLMKB02}. The strong coupling limit, $D \to \infty$, of an infinite
array of globally coupled systems was analytically treated already in the
pioneering paper \cite{VdB+94}, cf. also \cite{G-O+96,GM99}.

In this paper we investigate nonequilibrium phase transitions in {\it finite}
arrays of globally coupled Stratonovich models in the strong coupling limit.
We introduce center of mass and relative coordinates and exploit that for
strong coupling there is a clear separation of time scales.  The relative
coordinates relax very quickly to zero and the system behaves as a single
entity described by the center of mass coordinate $R_t$. Thus, we can
adiabatically eliminate the relative coordinates. The stationary probability
density of the center of mass coordinate $p_s(R)$ is analytically calculated
for a class of nonlinear systems and a scheme to determine the transition
manifold in the parameter space is developed. For finite arrays of
Stratonovich models the mean value $ \langle R \rangle$ of the center of mass
coordinate is computed analytically. Near a critical value of the control
parameter $a$ the stochastic system shows a scaling behaviour similar to the order parameter of
the single Stratonovich model with the same critical exponent $\beta=1$ but
with a different $a_c(N)$ which is also given analytically. Keeping a finite
small distance to $a_c(N)$ we recover for $N \to \infty$ the known result of
the self consistent theory \cite{BLMKB02} with critical exponent $\beta=1/2$,
see above. For finite $N$ we identify a crossover value of the control
parameter $a_\star(N)$. For $a_c(N)< a \ll a_\star(N)$ we have a linear
scaling as for $N=1$ whereas for $a \gg a_\star(N)$ a square root behaviour as
for $N \to \infty$ is observed. The analytical results are coroborrated by
numerical simulations.

Recently, finite arrays of (non-) linear stochastic systems have been 
investigated also by Mu\~noz et al.~\cite{MCC05}, and by Hasegawa
\cite{Hasegawa06a}. 

Mu\~noz et al.~tried to obtain for multiplicative noise characteristics of the
probability density of the mean field for finite $N$. They argued that the
Langevin equation  for the mean field variable is of similar form as the
Langevin equation for a single system. Assuming that the multiplicative
driving noise and the local field variable are uncorrelated, they inferred the
scaling behaviour of the variance of an effective multiplicative noise with
$N$, and of the critical value $a_c(N)$ of the control parameter. They also
predicted a crossover from a critical exponent $\beta=1$ near  $a_c(N)$ to the
critical exponents for $N \to \infty$ for larger distances to $a_c(N)$. Note
that in \cite{MCC05} the Langevin equation was treated in the Ito-sense which
leads to a shift of the critical control parameter compared to the same equation
in the Stratonovich-sense.

Hasegawa considered finite systems with additive and multiplicative noise
using his augmented moment method which is applicable for small noise
strength. He emphasized that multiplicative noise and the local field
variable are not uncorrelated in contrast to the assumption in \cite{MCC05}
and demonstrated some consequences of such a simplification.

Our approach, though similar in spirit to \cite{MCC05},  is controllable, 
valid in  leading order for strong coupling $D$, and provides  explicit
analytic results which are confirmed by independent numerical simulations.
It   may serve as a starting point to calculate next order corrections $\sim
1/D$. 

The paper is organized as follows. In the next section we consider two
harmonically coupled Stratonovich models and show that for strong coupling
the center of mass coordinate $R$ is the relevant degree of freedom. The mean
value of $R$ shows a critical behavior which is analytically characterized.
Section \ref{arbN} deals with a class of $N$ globally coupled systems of
general kind. For strong coupling we compute analytically the stationary
probability distribution $p_s(R)$ after eliminating the relative
coordinates. Further, we  determine the transition manifold in the parameter
space where $p_s(R)$ undergoes a transition from a delta-distribution to a
spatially extended solution. In Sec. \ref{NSM} we specialize to the case
of $N$ globally coupled Stratonovich models and determine the critical
behaviour of the order parameter and of higher moments of $R$ for strong
coupling. 
Conclusions are drawn and a summary is given in Section \ref{Con}.
In \ref{WeakNorm} we introduce the concept of weak
normalization for the case that a spatially extended solution of the
stationary FPE cannot be normalized in the naive sense. 
\ref{StratIto} shows that the Langevin approach both in  Stratonovich- and in
Ito-interpretation   leads to the same results as the Fokker-Planck approach
used in the main part of the paper.


\section{Two coupled  Stratonovich systems}\label{Neq2}

We consider a pair of particles with coordinates $x_1(t)$ and $x_2(t)$ in a
biquadratic  potential which are coupled harmonically with positive coupling
strength $D$ and each subjected to independent Gaussian white noise of
strength $\sigma$. The  system of Langevin equations reads 
\begin{eqnarray}
dx_i=\Big[ax_i-x_i^3- 
D\sum_{j=1,2}\big(x_i-x_j\big)\Big]dt 
+\sigma x_i\circ dW_i(t), \;\; i=1,2 \;,   \label{2LE}
\end{eqnarray}
where $W_i(t)$ denotes independent Wiener processes with  
$\langle W_i(t) W_j(s) \rangle = \delta_{i,j} \min{(t,s)}$.    
In contrast to  Eq. (\ref{1LE}) no exact solution of system (\ref{2LE}) is known.

The joint probability density $P(x_1,x_2;t)$ is governed by the FPE
\begin{eqnarray}
\partial_t P = - \sum_{i=1,2} \partial_{x_i} \Big[\big( D_i - \sum_{j=1,2}
\partial_{x_j} D_{i,j}\big) P\Big] ,\label{2FPE}
\end{eqnarray}
where, adopting the notation of \cite{Risken96},
\begin{eqnarray}
D_i(x_1,x_2)
=\big(a+\frac{\sigma^2}{2}\big)x_i-x_i^3-D\sum_{j=1}^2
(x_i-x_j),\\
D_{i,j}(x_i)=\frac{\sigma^2}{2} x_i^2 \delta_{ij}
\end{eqnarray}
denote drift and diffusion coefficients, respectively. 

One can show that the system (\ref{2FPE}) exhibits  no detailed balance.
Hence, there is no easy way to obtain analytically the stationary solution
$P_s$. 

\begin{figure}[h]
\begin{center}
	\includegraphics[angle=0, width=\textwidth]{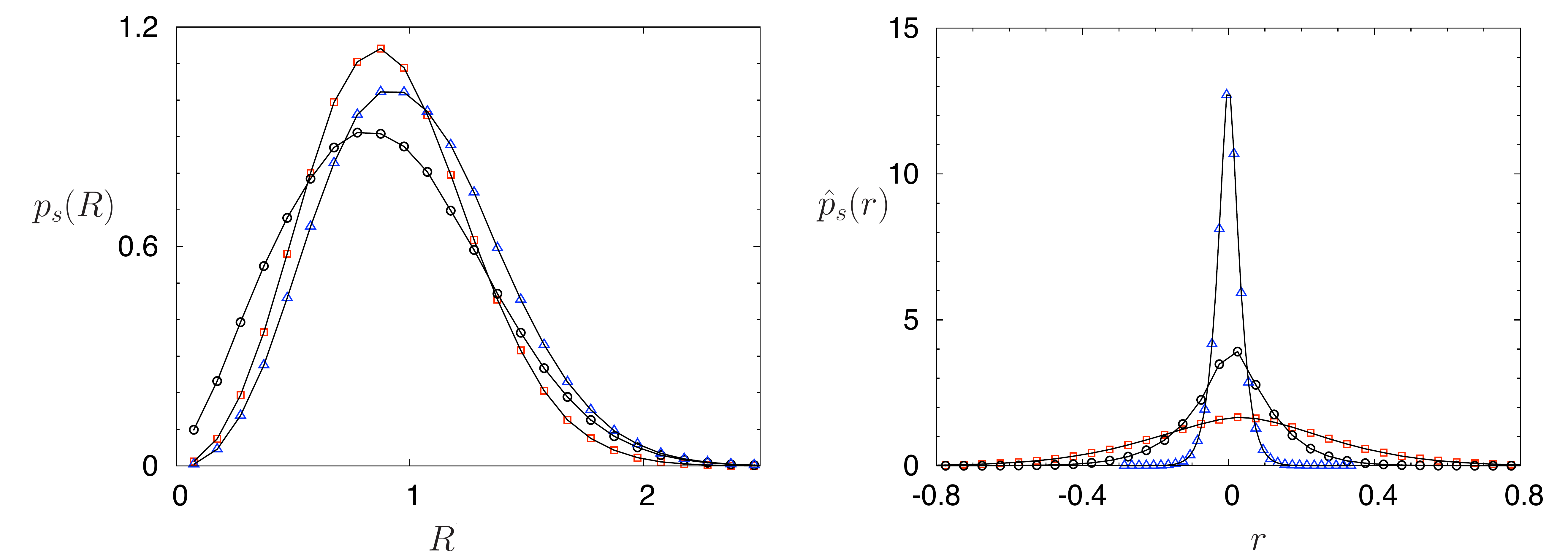}
\end{center}
\caption{
Stationary probability densities for center of mass $p_s(R)$ (left) and relative
coordinates $\hat{p}_s(r)$ (right) for two coupled systems. 
The distribution of relative coordinates is symmetric with respect to zero and becomes very
sharp with increasing strength of the coupling $D$. 
The symbols show histograms from $4\times 10^5$ realizations obtained by 
solving Eq. (\ref{2LE}) with a stochastic Runge-Kutta scheme \cite{SODESim}. 
Parameters are $D=1$ (squares),
$10$ (circles), and $100$ (triangles); $a=1$ and $\sigma^{2}=1$.  
Initial values were all in the positive sector. 
Entries of several bins are omitted to avoid overloading; the lines are guides to the eye.
}
\label{fig:R_r-hist_D}
\end{figure}

For strong coupling, however, a systematic analytical approach is possible.
With increasing coupling strength the particles become tightly glued together
and move as a single entity. Therefore it appears natural to introduce center
of mass and relative coordinates. Simulations of Eq. (\ref{2LE}) show that
indeed the stationary distribution of the relative coordinate 
$\hat p_{s}(r)$ becomes
very sharp for large values of $D$, cf. Fig. \ref{fig:R_r-hist_D}. 

\begin{figure}[h,b]
\begin{center}
	\includegraphics[angle=0, width=\textwidth]{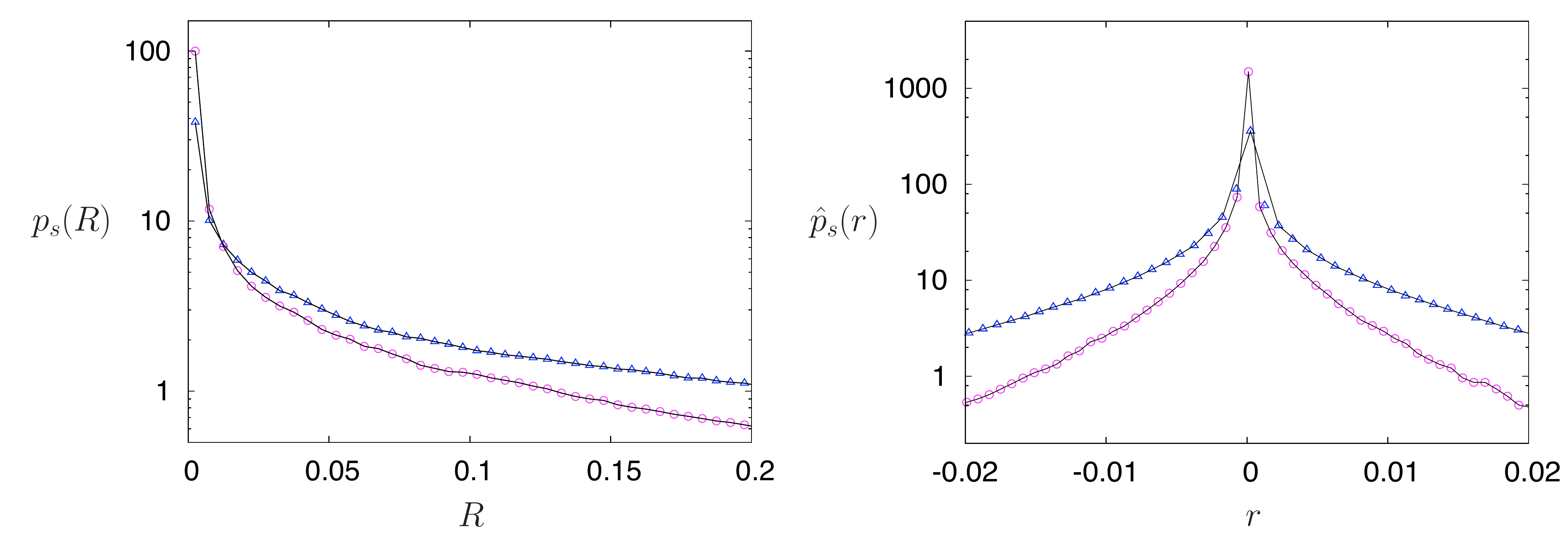}
\end{center}
\caption{
Stationary probability density for center of mass $p_s(R)$ (left) and relative
coordinates $\hat{p}_s(r)$  (right) for two coupled systems in a
semilogarithmic plot. For sufficiently small control parameter, $p_s(R)$
diverges for $R \to 0$ in a normalizable way. $\hat{p}_s(r)$ becomes
essentially sharper compared to   Fig.~\ref{fig:R_r-hist_D}. Parameters are 
$a=-0.05$,  $D=100$,   $\sigma^{2}=1$ (triangles) and $0.49$ (circles). The
symbols show data from $5\times 10^6$ realizations generated by a stochastic
Runge-Kutta algorithm \cite{SODESim}. Initial values were chosen in the positive
sector. The lines are guides to the eye.
} 
\label{fig:R_r-hist_a}
\end{figure}

The stationary distribution of the center of mass $p_{s}(R)$ shows a 
behaviour which is similar to  the distribution of a single Stratonovich
model. For large values of $a$ we have a monomodal distribution which vanishes
at the boundaries of the support, cf. Fig. \ref{fig:R_r-hist_D}. For small
values of $a$ the distribution $p_{s}(R)$  diverges as $R \to 0$ in a
normalizable way, cf. Fig. \ref{fig:R_r-hist_a}. For even smaller values of
$a$ all trajectories $x_i(t)$ approach zero and the distributions of both $r$
and $R$ are $\delta$-distributions. Accordingly, the mean value $\langle R
\rangle$ undergoes a continuous transition at a critical value of $a$. In the
following we analytically calculate $p_{s}(R)$ and $\langle R \rangle$ and
its  scaling characteristics  in the strong coupling limit, $D \to \infty$. 
We introduce the center of mass coordinate
$R(t)$  and the relative coordinate $r(t)$ by
\begin{eqnarray}
R=\frac{1}{2}(x_1+x_2), \;\; r=\frac{1}{2}(x_1-x_2)
\end{eqnarray}
with the inverse transformation
\begin{eqnarray}
x_1=R+r, \;\; x_2=R-r.
\end{eqnarray}
With
\begin{eqnarray}
\partial_{x_{1/2}}&=\frac{1}{2}\big(\partial _R\pm\partial_r\big),\\
\partial^2_{x_{1/2}}&=\frac{1}{4}\big(\partial^2_R + \partial^2_r \pm\partial^2_{Rr}\big),
\end{eqnarray}
the Langevin equations (\ref{2LE}) then transform to
\begin{eqnarray}
dR&=\left(aR-R^3-3Rr^2\right)dt 
+\frac{\sigma}{\sqrt{2}}\label{2R}
\left(R\circ d \widetilde W_1(t)+r\circ d\widetilde W_2(t)\right),
\\
dr&=\left[\left(a\!-\!2D\right)r\!-\!r^3\!-\!3rR^2\right]dt
+\frac{\sigma}{\sqrt{2}}\left(r\circ d\widetilde W_1(t)\label{2r}\!
+\!R\circ d\widetilde W_2(t)\right),
\end{eqnarray}
where the transformed Wiener processes $\widetilde W_i(t)$ are  defined as
\begin{equation}
\left(
\begin{array}{c}
\widetilde{W}_1\\
\widetilde{W}_2
\end{array}
\right)
=\frac{1}{\sqrt{2}}
\left(
\begin{array}{cc}
  1 &1 \\
  1 &-1
\end{array}
\right)
\left(
\begin{array}{c}
W_1\\
W_2
\end{array}
\right)
\end{equation}
with  
$\langle \widetilde W_i(t) \widetilde W_j(s) \rangle = \delta_{i,j} 
\min{(t,s)}$.   

The  FPE associated to (\ref{2R},\ref{2r}) governing the probability density 
of center of mass and relative coordinates $P(R,r;t)$ reads
\begin{eqnarray}
\partial_t P= \left({\cal L}_R+{\cal L}_r + {\cal L}_{rR}  \right)P ,  
\label{2FPE_R}
\end{eqnarray}
where the Fokker-Planck operators are 
\begin{eqnarray}
{\cal L}_R&=&-\partial_R
  \Big[\big(a+\frac{\sigma^2}{2}\big)R-R^3-3Rr^2\!-\!\frac{\sigma^2}{4}\partial_R\left(R^2+r^2\right)\Big] ,\\
 {\cal L}_r&=&-\partial_r
  \Big[\big(a-2D+\frac{\sigma^2}{2}\big)r-r^3-3rR^2\!-\!\frac{\sigma^2}{4}\partial_r\left(R^2+r^2\right)
     \Big],\\ 
 {\cal L}_{r R}&=& \;\sigma^2 \partial^2_{rR} \; rR\;.
\end{eqnarray}
Note that only ${\cal L}_r$ depends on $D$. In the strong coupling limit
$D\to\infty$ the relative  coordinate vanishes, $r_t\to0$, on a very fast time
scale of the order $1/D$, cf. Eq. (\ref{2r}). Hence, the stationary 
probability density factorizes  to $P_s(R,r)=p_s(R)\delta(r)$ with a Dirac
distribution for the relative coordinate. In this case there is no flow
related to the relative coordinate $r$, i.e. ${\cal L}_r P ={\cal L}_{rR}
P\equiv 0$, since for any suitable function $\varphi$
\begin{eqnarray}
\int_{-\infty}^\infty dr \partial_r [\varphi(r) \delta (r)] \equiv 0\;.
\end{eqnarray}

Integrating  Eq. (\ref{2FPE_R}) with respect to $r$ yields in the stationary
case
\begin{eqnarray}
0 = \!\!\int\limits_{-\infty}^{\infty} \!\!dr {\cal L}_R P_s =-\partial_R
\Big[\big(aR-R^3-\frac{\sigma^2}{4}R^2\partial_R\big)p_s\Big]\,.\label{2dFP_red}
\end{eqnarray}
Similarly as for the single Stratonovich model, there is always a weak
solution $\delta(R)$. 

For initial values $x_i(0)>0 \;\; \forall i$ (or $x_i(0)<0 \;\;\forall i$) the
spatially extended solution of Eq. (\ref{2dFP_red}) lives on the support
$S_+=[0,\infty)$ (or on $S_-=(-\infty,0]$) and  can be normalized provided
$a>a_c(2)=-\sigma^2/4$. For $a\leq a_c(2)$ the weakly normalized version of
the spatially extended solution converges to $\delta (R)$. 
Thus, we have
\begin{eqnarray}\label{Ps2}
p_s(&R&)= 
\cases
{
    \delta(R) \hspace{2.75cm} \text{for}\,\,a \leq a_c(2),\\
    \frac{1}{Z}|R|^{4a/\sigma^2}e^{-2R^2/\sigma^2} \hspace{0.25cm} \text{for}\,\,a > a_c(2),
}\\
&Z&= \frac{1}{2} \left(\sigma/2\right)^{2a/ \sigma^2+1/2}\Gamma \left({2a}/{\sigma^2}+1/2\right).
\end{eqnarray}
There is a strong ergodicity breaking when $a$ crosses $a_c(2)$.
The mean value $\langle R \rangle _\pm$ calculated with (\ref{Ps2}) is
\begin{eqnarray}
\langle R \rangle _\pm = 
\cases
{
0 \hspace{4cm}\text{if} \;a \leq a_c(2),\\
\pm \frac{\sigma}{\sqrt{2}}
\frac{\Gamma({2a}/{\sigma^2}+{1})}{\Gamma({2a}/{\sigma^2}+1/2)}\hspace{0.4cm} \text{if} \;a > a_c(2).
}
\end{eqnarray}
and scales like $\langle R \rangle_\pm \sim \pm\frac{\sqrt{2}\pi}{\sigma} \big( a-
a_c(2)\big)^\beta $ with $\beta=1$ as $a \to a_c(2)$ .

\section{General N-site systems}\label{arbN}

\subsection{Adiabatic elimination of relative coordinates}

In the following we demonstrate that the strategy sketched above can be generalized for
a class of $N$ coupled systems. We consider 
\begin{eqnarray}
dx_i=\Big[f(x_i)
-\frac{D}{N-1}\sum_{j=1}^N\left(x_i-x_j\right)\Big]dt + g(x_i)\circ dW_i(t),\label{NdLE}
\end{eqnarray}
with $ i=1,\dots,N$ and where $f$ and $g$ are smooth (with no singularities) and twice differentiable  chosen such that the stochastic process 
$\mathbf{x}(t) = \{x_i(t), i=1, \dots, N\}$ has natural boundaries at infinity \cite{HL84,Gichman}. Both $f$ and
$g$ may depend on a $d$-dimensional set of control parameters $\mathbf{a}$. $D>0$ is
the coupling strength of the harmonic attraction. Note that we have absorbed a factor
$\sigma $, the strength of the noise, in the function $g$.

The FPE  for the joint probability density 
$P(\mathbf{x};t)$, $\mathbf{x} = \{x_i, i=1, \dots, N\}$, associated to (\ref{NdLE})
reads
\begin{eqnarray}
\partial_t P & =-\sum_{i=1}^N\partial_{x_i}
\Big[\Big(D_{i}-\sum_{j=1}^N\partial_{x_j} D_{i,j}\Big)P\Big].
\end{eqnarray}
Using the shorthands  $f_i=f(x_i)$,
$g_i=g(x_i)$, and $g_i'=\partial_{x_i} g_i$, 
drift coefficient and diffusion matrix are given by 
\begin{eqnarray}
D_{i}  &=& f_i +\frac{1}{2}g_i'g_i
-\frac{D}{N-1}\sum_{j=1}^N\left(x_i-x_j\right), \\
D_{i,j}  &=&\frac{1}{2}g_i^2\,\delta_{ij}.
\end{eqnarray}

It is advantageous to introduce center of mass
and relative coordinates $\{R, \mathbf{r}\}$, $\mathbf{r}=\{r_k,
k=2,\ldots,N\}$, by the  linear transformations 
\begin{eqnarray}
R &=\frac{1}{N}\sum_{i=1}^N x_i, \\ 
r_k &=x_k - R \qquad \text{for} \qquad k=2,\ldots,N \;.
\end{eqnarray}
The inverse transformation is given by 
\begin{eqnarray}
x_1 &=R-\sum_{k=2}^Nr_k \label{inversex1},\\
x_k &= R+r_k\ \qquad \text{for} \qquad k=2,\ldots,N\;.\label{inversexk}
\end{eqnarray}
Observing the rules for linear transformations we have 
\begin{eqnarray}
\sum_{i=1}^N\!\frac {\partial}{\partial x_i}D_i =
\frac {\partial}{\partial R} D_R + \sum_{k=2}^N\!\frac {\partial}{\partial r_k}
D_{r_k},\\
\sum_{i,j=1}^N\!\frac {\partial^2}{\partial x_i\partial x_j}D_{ij} = \frac {\partial^2}{\partial R^2} D_{R,R}  +
\sum_{k=2}^N\!\frac {\partial^2}{\partial R\partial r_k}D_{R,r_k}+
\sum_{k,l=2}^N\!\frac {\partial^2}{\partial r_k \partial r_l}D_{r_k,r_l}.
\end{eqnarray}
Drift and diffusion coefficients in the new coordinates are given by, cf. also
\cite{Risken96},
\begin{eqnarray}
D_{y}=\sum_{i=1}^N \frac{\partial y}{\partial x_i} D_{i}
\; , \quad
D_{y,z}=
\sum_{i,j=1}^N \frac{\partial y}{\partial x_i}
\frac{\partial z}{\partial x_j} D_{i,j},
\end{eqnarray}
where $y$ and $z$ stand for the new coordinates $R$, $r_k$, and $r_l$, respectively.

Again, the FPE determining  $P(R,\mathbf{r};t)$
has the form $\partial_t P={\cal L} P$, with  ${\cal L}={\cal L}_R+{\cal L}_r +
{\cal L}_{rR}$ 
where
\begin{eqnarray}
\label{FPinRr}
{\cal L}_R= -\partial_R \big( D_R -\partial_R D_{R,R} \big),\\
{\cal L}_r=-\sum_{k=2}^N \partial_{r_k} \big( D_{r_k}-
\sum_{l=2}^N \partial_{r_l}D_{r_k,r_l}\big),\\
{\cal L}_{rR}=\sum_{k=2}^N \partial^2_{Rr_k} D_{R,r_k}.
\end{eqnarray}
Explicitly, the new drift and diffusion coefficients are 
\begin{eqnarray}
D_R & =\frac{1}{N}\sum_{i=1}^N\Big(f_i+\frac{1}{2}g_i'g_i\Big)\,,
\label{DR}\\
D_{r_k}&=-D_R+f_k+\frac{1}{2}g_k'g_k-D\frac{N}{N-1}r_k\,,\\
D_{R,R} & = \frac{1}{2N^2}\sum_{i=1}^Ng_i^2\,,\label{DRR}\\
D_{R,r_k} & =D_{r_k,R}=\frac{1}{2N}g_k^2-D_{R,R}\,,\label{DRrk}\\
D_{r_k,r_l}&=D_{R,R}-\frac{1}{2N}\left(g_k^2+g_l^2\right)+\frac{1}{2}g_k^2\,\delta_{kl}\,.
\end{eqnarray}
All arguments in $f_i$, $g_i$, and  $g'_i$ have to be expressed by
$(R, {\mathbf r})$, see Eqs. (\ref{inversex1},\ref{inversexk}). Note that only
$D_{r_k}$ depends  on the coupling strength $D$ explicitly.

For large times the probability density $P(R,\mathbf{r};t)$ converges to a 
stationary probability density, cf. \cite{SB_Htheorem},
determined  by  $ {\cal L}P_s(R,\mathbf{r}) = 0$. For $D\to\infty$ this
enforces  
\begin{equation}\label{FPElargeD} 
\sum_{k=2}^N \frac {\partial}{\partial r_k} \big[r_k P_s(R,\mathbf{r})\big]=0\;, 
\end{equation} 
which has a weak solution  
\begin{equation}\label{weaksol}
P_s(R,\mathbf{r})=p_s(R)\delta(\mathbf{r}).   
\end{equation} 
In the strong coupling limit all fluctuations of the relative coordinates
vanish. The system is concentrated on the center of mass and moves 
stochastically as a whole, combined particle.

The probability density of the center of mass $p_s(R)$ can be determined 
by integrating $P_s(R,\mathbf{r})$ over all relative coordinates.
Performing this integration we obtain from the stationary FPE 
\begin{eqnarray}\label{FPint}
&\int d^{N-1}\mathbf{r}\;{\cal L}P_s
=\int d^{N-1}\mathbf{r}\; {\cal L}_R(R,\mathbf{r}) P_s(R,\mathbf{r})=0,
\end{eqnarray}
provided that the boundary terms associated with the relative coordinates
vanish. In the strong coupling limit we have $P_s(R,\mathbf{r}) \propto
\delta(\mathbf{r})$, (\ref{FPint}) holds in any case and leads to  
\begin{equation}\label{FPR}
{\cal L}_R(R,\mathbf{0}) p_s(R)=0 \;,
\end{equation}
where ${\cal L}_R$ is given by (\ref{FPinRr}). From (\ref{DR}) and (\ref{DRR})
we infer drift and diffusion for  $\mathbf{r}=0$ as
\begin{eqnarray}
 D_R(R,\mathbf{0})=f(R)\label{DriftcoeffN}
+\frac{1}{2} g'(R) g(R)\;,\\
D_{R,R}(R,\mathbf{0})  = 
\frac{1}{2N} g^2(R)\,.\label{DiffcoeffN}
\end{eqnarray}

The spatially extended strong solution of (\ref{FPR}) is given by
\begin{equation}\label{Ps}
p_s(R)=\frac{1}{Z}\;|g(R)|^{N-2}\; \exp \Big\{ 2N\!\int^R dR' 
\!\frac {f(R')}{g^2(R')} \Big\} 
\end{equation}
provided that the normalization constant $Z$ is finite. Whether or not this is
the case depends on the  scaling behaviour of the functions $f(R)$ and  $g(R)$
near a common zero $R_0$ which, if existing, will build a boundary of the
support. This is explained in detail in the next subsection.

Equation (\ref{DiffcoeffN}) shows that in the strong coupling  limit the
diffusion coefficient $D_{R,R}$ scales like  $\sigma^2/N$, cf. also
\cite{MCC05,Hasegawa06a}. For the infinite system and finite noise strength
$\sigma$ the stationary probability density of the center of mass $p_s(R)$ is
a Dirac measure located at one of the attractive zeros of the drift
coefficient (\ref{DriftcoeffN}), depending on the initial conditions.

For $D \to \infty$ all particles are strongly correlated. The variance of the
coordinate $x_i(R,\mathbf{r})$ of an arbitrary system $i$ calculated with 
$P_s(R,\mathbf{r})=p_s(R)\delta(\mathbf{r})$ is
\begin{eqnarray}
\langle x_i^2\rangle-\langle x_i\rangle^2
=\langle R^2\rangle-\langle R\rangle^2.
\end{eqnarray}
Due to the strong correlations,  the variance of the center of mass scales
like $N^0$ in contrast to the case of uncorrelated systems where the central
limit theorem predicts a scaling  like $N^{-1}$.

\subsection{Determination of the transition manifold}

There will be a strong ergodicity breaking if the state space decomposes into
different regions with the property that one region will not be accessible if
we start in a different one \cite{BB06}. 

For multiplicative noise, zeros of the stochastic flow separate the state
space into mutually non-accessible regions. If we place the system initially
on such a zero, i.e. on a fixed point of the stochastic dynamics, it will
stay there forever. Accordingly, the FPE has a weak solution, a
$\delta$-distribution living on that fixed point.  If any trajectory in the
neighborhood 'asymptotically' reaches the fixed point (the fixed point is
absorbing), there will be no spatially extended probability density in this
neighborhood. The spatially extended stationary solution of the FPE cannot
be normalized in the naive sense. The weak normalization procedure leads to
the weak solution. 

If trajectories cannot reach the fixed point,  the stationary solution of the
FPE will be normalizable and we will have a spatially extended probability
density living on the support bounded by the fixed point. This properties can
be exploited to determine the transition manifold in the parameter space.

We suppose that $f(R,\mathbf{a})$ and $g(R,\mathbf{a})$ near a common zero
$R_0$ have the following scaling behaviour  
\begin{eqnarray}
f(R_0+\varepsilon)&\sim A_f \varepsilon^{m_f},\label{fscaling}\\
g(R_0+\varepsilon)&\sim A_g \varepsilon^{m_g},
\end{eqnarray}
where $m_f, m_g >0$.
Near $R_0$ we have for the stationary solution (\ref{Ps}) of the reduced FPE
(\ref{FPR}) 
\begin{eqnarray}
p_s(R_0+\varepsilon) \propto  |\varepsilon|^{m_g(N-2)} \exp \Big\{ 2N \!\!\!\!
\int\limits^{R_0+\varepsilon}\!\!dR'
\frac {f(R')}{g^2(R')}\Big\} .   \label{P(eps)}     
\end{eqnarray}

For $ m_f-2m_g > -1 $ the integral in (\ref{P(eps)}) gives  a contribution
$\propto \varepsilon^{m_f-2m_g+1}$ at the upper boundary  which vanishes for
$\varepsilon  \to 0$ so that in leading order  $p_s(R_0+\varepsilon) \propto 
|\varepsilon|^{m_g(N-2)}$. The exponent $m_g(N-2)$ is negative only for $N=1$.
In this case, if $m_g \geqslant 1$ the singularity of $p_s$ at $R=R_0$ is not
normalizable and we have only a weak stationary solution
$p_s(R)=\delta(R-R_0)$. Note that for $N \geqslant 2$ coupled systems of this
kind the singularity of $p_s$ does not occur.

For $ m_f-2m_g = -1 $ the integral gives a logarithmic contribution
$(A_f/A^2_g)\ln|\varepsilon|$ which leads to 
\begin{eqnarray}
p_s(R_0+\varepsilon) \propto \; |\varepsilon|^{m_g(N-2) +2NA_f/A^2_g}\;. 
\label{P(eps)2}
\end{eqnarray}
If the exponent in (\ref{P(eps)2}) is smaller than $-1$ the density  $p_s(R)$
will diverge for $R \to R_0$  and will not be normalizable in a naive way. The
weak normalization procedure leads to a Dirac measure located at $R_0$. If the
exponent is larger than  $-1$ the density $p_s(R)$ will be normalizable and we
will have a spatially extended probability density.  The transition manifold
${\cal A}_c$ in the control parameter space $\mathbb R^d$  is determined by
the condition that the exponent in (\ref{P(eps)2}) is $-1$,
\begin{eqnarray}
{\cal A}_c&=\left\{\mathbf{a}\in\mathbb R^d: m_g(N-2) +2NA_f/A^2_g =-1
\right\}\label{mani}\,.
\end{eqnarray}

For $m_f-2m_g<-1$ the integral in (\ref{P(eps)}) diverges like $
-A_f/(A_g^2|m_f-2m_g+1|) \varepsilon^{-|m_f-2m_g+1|}$ as $\varepsilon \to 0$.
Accordingly, $p_s(R \to R_0) = 0$ for $A_f >0$ and $p_s(R \to R_0) = \infty$ 
for $A_f <0$. In the first case $p_s(R)$ is normalizable and we have a spatially
extended stationary probability density. In the latter case the weak normalization
procedure yields a Dirac measure at $R_0$. A change in the sign of $A_f$ induced by
tuning a control parameter is associated with a change of the stability of the fixed
point $R_0$ of the {\it deterministic} flow $f(R)$ and leads to a significant
alteration of the ergodic properties. In a vicinity of $R_0$ the behaviour of the
stochastic system is dominated by the deterministic flow. The transition manifold 
\begin{eqnarray}
{\cal A}_c&=\left\{\mathbf{a}\in\mathbb R^d:A_f=0 \right\}\label{A_fzero}
\end{eqnarray}
does not depend on the system size and the amplitude $A_g$ in contrast to
(\ref{mani}).
If the system lives on the $d-1$ dimensional transition manifold (\ref{A_fzero}),
that is $A_f=0$, the scaling of the deterministic flow is not described by
(\ref{fscaling}) but by  $f(R_0+\varepsilon) \sim B_f \varepsilon^{n_f}$ with $n_f >
m_f$. The systems behaviour, now depending on $B_f, A_g, n_f, m_g$ and $N$, could be
classified in more detail repeating the above procedure.

\section{$N$ coupled Stratonovich  models}\label{NSM}
Now we return to our specific example and consider $N$ globally 
coupled Stratonovich models in the strong coupling limit. 
For drift and noise function we have
\begin{eqnarray}
f(R;a)=aR-R^3\quad\text{and}\quad g(R;\sigma)=\sigma R\,.
\end{eqnarray}
The common zero of $f$ and $g$ is $R_0=0$ with $m_f-2m_g=-1$,  and
$A_f=a$, $A_g=\sigma$. Inserting this in (\ref{mani}) we obtain an explicit
representation of the transition curve in the 2-dimensional parameter space,
\begin{eqnarray}
{\cal A}_c=
\left\{(a,\sigma)\in\mathbb R^2: N-1+2Na/\sigma^2=0
\right\}.
\end{eqnarray}
Given the noise strength $\sigma$ we have
\begin{eqnarray}\label{ac}
a_c(N)=-\frac{\sigma^2}{2}\Big(1-\frac{1}{N}\Big)\;,
\end{eqnarray}
which reproduces the results for $N=1$, for $N=2$ (see above), and for $N \to
\infty$.
\begin{figure}[h,b]
\begin{center}
  \includegraphics[width=0.5\textwidth, angle=0]{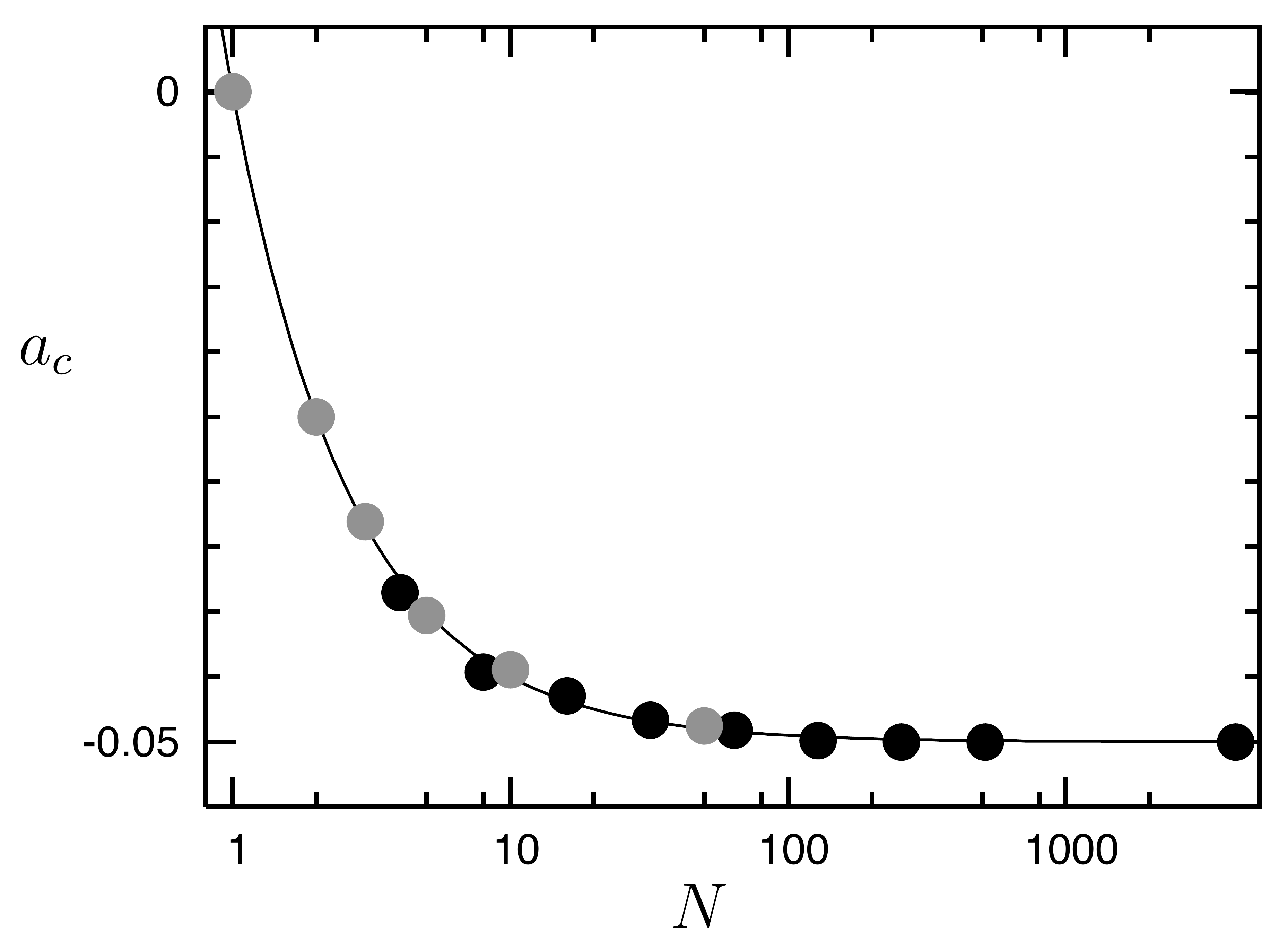}
\caption{\label{ac_vs_N} Transition point $a_c$ vs system size $N$ for 
$\sigma^2=0.1$.   Simulation results for  $D=81$ obtained 
by maximizing the linear  correlation coefficient \cite{MaxCorr} (gray bullets) and
by short time relaxation \cite{ShortTimeRelax} (black bullets) are both in good
agreement  with the asymptotic result (\ref{ac}) for $D\to \infty$ (solid line).}
 \end{center} 
\end{figure}
Figure~\ref{ac_vs_N} compares results from simulation 
and the asymptotic result (\ref{ac}) for $a_c(N)$ and illustrates  the finite
size scaling $a_c(N)-a_c(\infty)=\sigma^2/(2N)$ for strong coupling
$D\gg1$.

For  initial values $x_i >0$ (or $x_i<0$) $\forall i$ the stationary
distribution for the center of mass (\ref{Ps}) lives on $S_+$ or $S_-$,
respectively, and is given by
\begin{eqnarray}
\label{PsSM}
p_s(R)&=
\cases
{ 
\delta(R) \hspace{4cm}\text{if} \;\;a \leq a_c(N),\\
\frac{1}{Z}|R|^{\frac{2N}{\sigma^2}(a-a_c(N))-1}e^{-\frac{N}{\sigma^2}R^2}\hspace{0.45cm}\text{if} \;\; a > a_c(N),
}\\
Z &=\frac{1}{2} ({\sigma^2}/{N})^{\frac{N}{\sigma^2}\left(a-a_c(N)\right)}
  \Gamma\big((a-a_c(N))N/\sigma^2\big)\;.
\end{eqnarray}
For initial values $x_i =0 \; \forall i$ we have  $p_s(R)=\delta(R)$ for all
values of $a$. 

\begin{figure}
\begin{center}
   \includegraphics[width=\textwidth, angle=0]{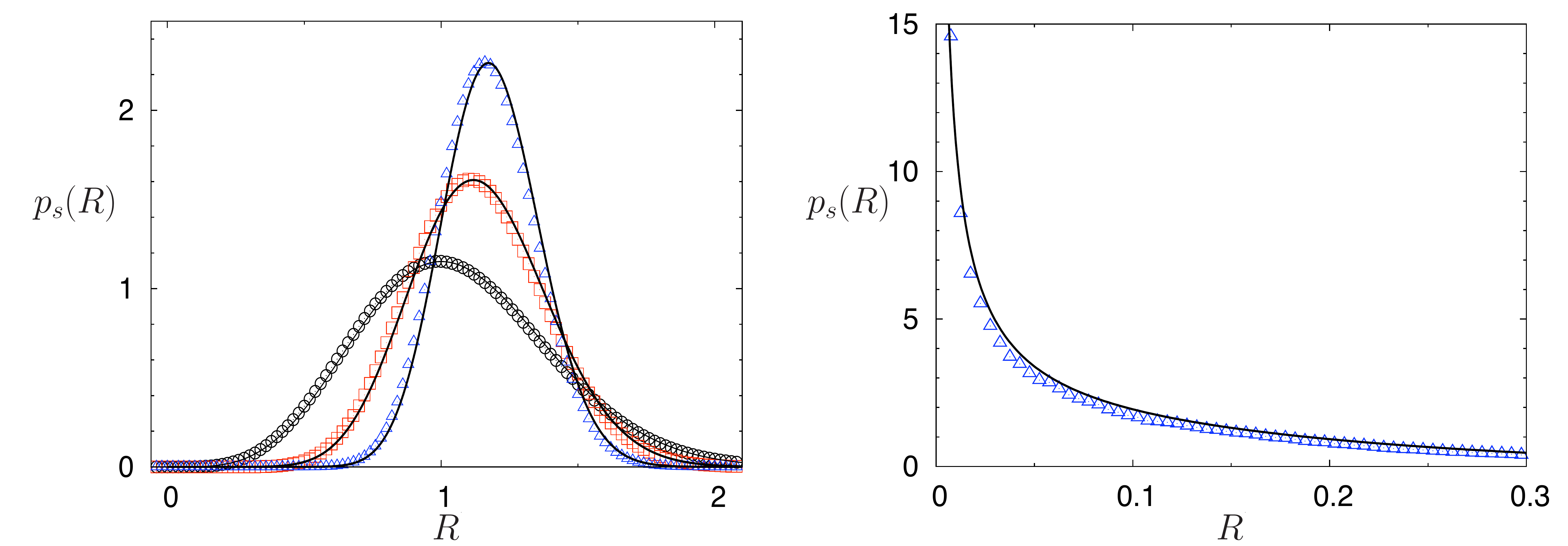}
\end{center}    
 \caption{\label{PvsR}
Probability density of the center of mass obtained by simulation for $D=100$
(symbols) compared with analytic results (lines) for $D\to \infty$ given by
Eq. (\ref{PsSM}). On the left we show results for different system sizes $N=2$
(circles), $4$ (squares), and $8$ (triangles)  for $a=1 > a_c^{\text{max}}$.
The histograms are obtained from $10^6$ realizations generated by a stochastic
Euler method \cite{EulerMaruyama}. On the right we show
only results for $N=8$ for $a_c < a=-0.42 < a_c^{\text{max}}$; 
here $5\times10^6$ realizations were generated by a stochastic Runge-Kutta algorithm 
\cite{SODESim}. $\sigma^2=1$.
}
\end{figure}

Similar to the single Stratonovich model, there is a qualitative change in
the shape of the spatially extended probability density. The maximum of
$p_s(R)$ undergoes a bifurcation at $a_c^{\text{max}}= \sigma^2(1/N-1/2)$.
Figure \ref{PvsR} compares for  different system sizes the asymptotic result
(\ref{PsSM}) with  histograms obtained by simulations  for large $D$. 
For $ a > a_c(N)$  the $n$th moment of the center of mass
can be evaluated as 
\begin{eqnarray}
\left\langle R^n\right\rangle_\pm
= (\pm)^n \Big(\frac{\sigma^2}{N}\Big)^{\!\textstyle\frac{n}{2}}\;\frac{\Gamma\big((a-a_c(N))N/\sigma^2 
+n/2\big)} {\Gamma\big((a-a_c(N)) N/\sigma^2\big)}\;.\label{R^n}
\end{eqnarray}
Keeping  $N$ finite, the first moment scales as $a \to a_c(N)$ like
\begin{eqnarray}
\left\langle R\right\rangle_\pm \sim \pm\frac {N^{1/2}}{\sigma} 
\sqrt{\pi} \big(a-a_c(N)\big)^{\beta},\;\beta=1,
\end{eqnarray}
since $\Gamma (z) \sim 1/z$ as $z \to 0$ \cite{Olver}. Note that also the higher
moments $\left\langle R^n\right\rangle$ scale linear with $a-a_c(N)$. 

Keeping a finite distance to $a_c(N)$ we obtain for $N \to \infty$, observing 
$\Gamma (z+ 1/2)/\Gamma(z)=\sqrt{z} (1-1/(8z)+ \dots)$ as $z \to \infty$
\cite{GKP94}, 
\begin{eqnarray}
\left\langle R\right\rangle_\pm \sim \pm\big(a-a_c(\infty)\big)^{\beta}, \; 
\beta=1/2,
\end{eqnarray}
which reproduces the result in \cite{BLMKB02} for $D > \sigma^2$. Higher
moments of order $n$ scale with $\beta =n/2$.

\begin{figure}[h]
\begin{center}
	\includegraphics[width=0.75\textwidth, angle=0]{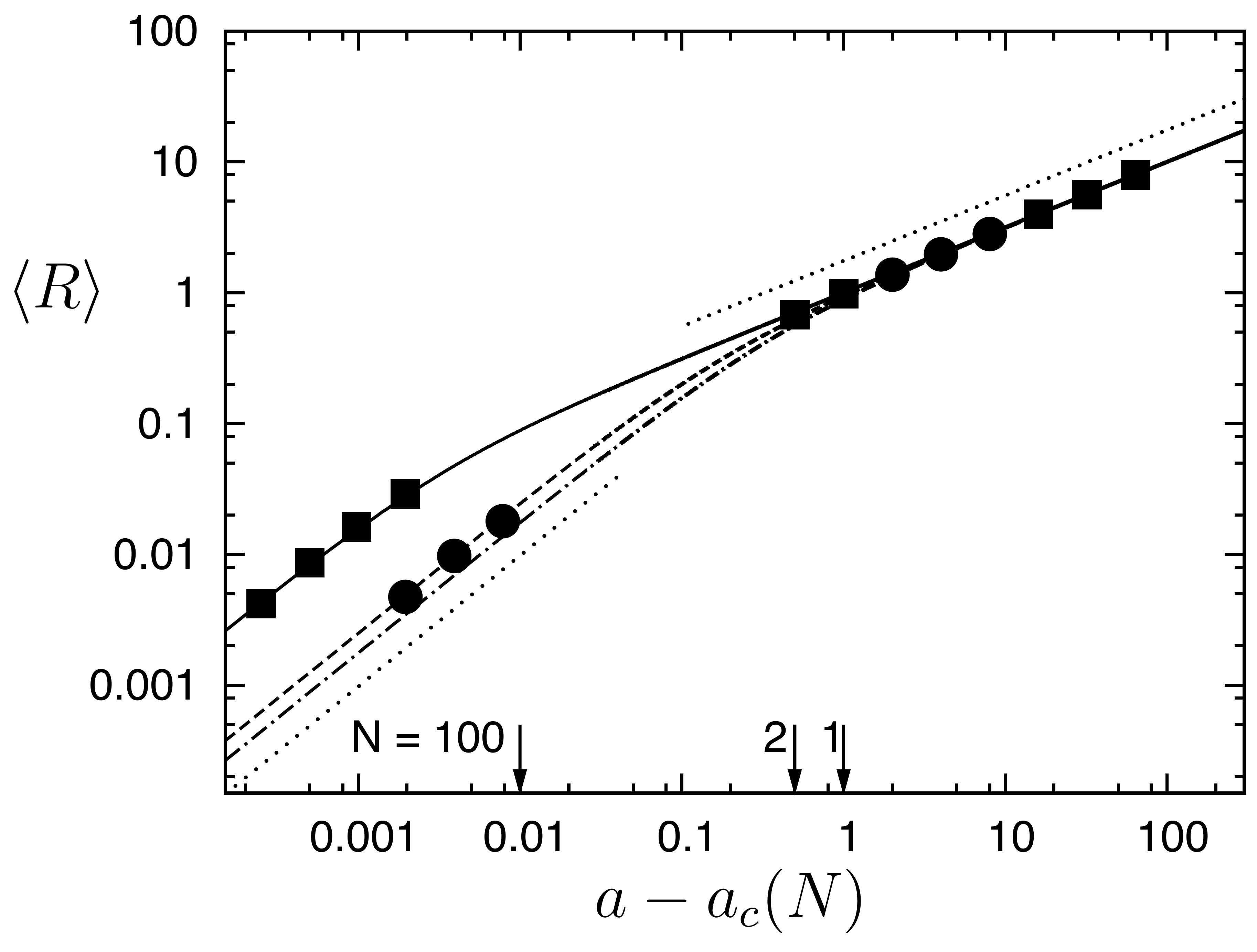}
\end{center}	
	\caption{\label{crossover}
Crossover in the scaling behaviour of $\langle R \rangle$ as a function of
$a-a_{c}(N)$ as predicted by Eq. (\ref{R^n}) for $N=1$ (dash-dotted line), 
$2$ (dashed line), and $100$ (solid line); $\sigma^{2}=1$.  The symbols show
averages over  $2\times10^6$  realizations generated by a stochastic
Runge-Kutta scheme \cite{SODESim} for $N=2$ (circles) and $100$  (squares);
$\sigma^{2}=1$ and $D=100$.  The dotted straight lines have  slope $1$ (left)
and slope $1/2$ (right), respectively. The arrows indicate the crossover
points $a_{\star}(N)-a_{c}(N)$.
}
\end{figure}

We define the crossover value $a_\star(N)$ by $(a_\star(N)-a_c(N))
N/\sigma^2=1$. For $a \ll a_\star(N)=-\sigma^2/2+(3/2)\sigma^2/N$ we have a
linear scaling as for $N=1$ whereas for $a \gg a_\star(N)$ a square root
behaviour as for $N \to \infty$ is observed, cf. Fig. \ref{crossover}.

Our results are analytically derived  for the strong coupling limit in a
controllable approach. We note that both the critical and the crossover value of the
control parameter are in accordance with the values proposed on different grounds in
\cite{MCC05} for weak and intermediate noise, provided the shift due to the Ito
interpretation used there is taken into account.

\section{Conclusions}\label{Con}

In this paper we have determined the characteristics of a continuous nonequilibrium
phase transition in a finite array of globally coupled Stratonovich models in the
limit of strong coupling $D \to \infty$. In this limit there is a clear separation
of the time scales governing the evolution of the center of mass coordinate and the
relative coordinates: The characteristic time of the relative coordinate scales as $
1/D$ and thus becomes short in the strong coupling limit. The slow center of mass
coordinate enslaves the fast relative coordinates, its mean value serves as order
parameter. This allows a controllable and consistent treatment both in the
Fokker-Planck and the Langevin description which is corroborated by numerical
simulations. 
The reduction of a high-dimensional problem to a problem of low dimension is of
course inspired by generalizations of slaving and adiabatic elimination techniques
and the concept of center manifolds to stochastic systems developed  in a different context
\cite{HW82,SH86,CR96}, cf. also \cite{Gardiner,Arnold}. 

We have analytically determined both the critical value of the control parameter
$a_c(N)$ and the scaling behaviour of the order parameter and of higher moments.
With increasing distance from $a_c(N)$ a crossover from linear to square root
behaviour is found. For $N \to \infty$ the known scaling behaviour is reproduced.
The formal results, i.e., the computation of the stationary distribution of the
center of mass coordinate (up to a quadrature) and the determination of the
transition manifold are given for a general class of systems.

Our approach may serve as a starting point to calculate next order corrections in
$1/D$. In a multiscale analysis we have to take into account  that for finite but
large $D$ the distribution of the relative coordinates is, though very sharp, of
finite width.

The observation that a solution of the stationary Fokker-Planck equation which
is not normalizable in a naive way converges to the weak solution if weakly
normalized is certainly of interest in a broader context.

\FloatBarrier

\begin{appendix}
\section{Weak normalization}\label{WeakNorm}

The FPE for multiplicative noise has two types of stationary solutions: weak
solutions, i.e. Dirac-distributions living on the zeros of the stochastic flow
and spatially extended strong solutions which live on a support which is
bounded by zeros of the stochastic flow or by natural boundaries at infinity.
Under certain conditions the spatially extended solution may diverge at a zero
of the stochastic flow so strongly that it is not normalizable and therefore
cannot be considered as a probability density. Here we introduce the concept
of weak normalization and show that in the latter  case the weakly normalized
solution converges to the Dirac distribution living on that zero.

We assume that the unnormalized solution $\tilde P_s(x)$ lives on $[x_0,b)$
where $x_0$ is a zero of the stochastic flow and scales for $x \to x_0$ as 
\begin{eqnarray}
\tilde P_s(x) \sim {\text {const}} \;(x-x_0)^{\alpha} , \;\; \alpha < -1\; .
\end{eqnarray}
The normalization integral diverges at the lower boundary and scales like 
\begin{eqnarray}
\int_{x_0+\Delta}^b dx \tilde P_s(x) \sim -{\text {const}}\;  \frac {1}{\alpha +1} \Delta^{\alpha
+1}\;\; {\text {as}} \;\;\Delta \to 0.
\end{eqnarray}
Introducing a test function $\varphi (x)$ which can be expanded near $x_0$ as
$\varphi(x)=\varphi(x_0) + \varphi'(x_0)(x-x_0)+ \dots$ we have as $ \Delta \to 0$
\begin{eqnarray}
\int_{x_0+\Delta}^b dx \tilde P_s(x) \varphi (x) \sim&& - {\text { const}} 
\frac{1}{\alpha +1} \Delta^{\alpha +1} \nonumber \\ 
&&\times \{\varphi(x_0) + \frac{\alpha +1}{\alpha +2}\varphi'(x_0)\Delta + \dots\}\;.
\end{eqnarray}
Now we obtain for the hereby defined weakly normalized probability density
$P_s^w(x)$
\begin{eqnarray}
\int_{x_0}^b dx  P_s^w(x) \varphi (x)= \lim_{\Delta \to 0}
\frac{\int_{x_0+\Delta}^b dx \tilde P_s(x) \varphi (x)}{\int_{x_0+\Delta}^b dx \widetilde
P_s(x)}= \varphi(x_0),
\end{eqnarray}
which implies that $P_s^w(x)=\delta(x-x_0)$.

\section{Langevin approach}\label{StratIto}

In Sections \ref{Neq2} and \ref{arbN} we used the Fokker-Planck approach in
the center of mass and relative coordinates to calculate $a_c(N)$ for $D \to
\infty$. In this limit the relative coordinates $r_k \to 0$, and it is easy to
calculate the reduced stationary probability density of the center of mass
coordinate $p_s(R)$. We determined $a_c(N)$ such that $p_s(R)$ is a Dirac
measure at $R=0$ for $a<a_c(N)$  and it is spatially extended for $a>a_c(N)$.

The same result can be obtained in the Langevin approach, both in Stratonovich
and Ito-interpretation as explained in the following for the special case $N=2$. The
generalization to $N>2$ is straightforward.

We exploit that for large $D$ the characteristic time scale of the relative
coordinate $r(t)$ is $1/D$ so that $r(t)$ becomes very fast. Then the
(slow) center of mass coordinate $R(t)$ feels only the average of the fast
process $r(t)$ and it is justified to replace in the Stratonovich-Langevin
equation (\ref{2R}) for $R$ the terms associated with $r$ by their average,
\begin{eqnarray}
dR=\left(aR-R^3-3R \langle r^2 \rangle \right)dt\!
+\!\frac{\sigma}{\sqrt{2}}
\left(R\circ d \widetilde W_1(t)+ \langle r\circ d\widetilde W_2(t)\rangle 
\right),
\end{eqnarray}
since for $D \to \infty$, $r \to 0$ we have $\langle r^2 \rangle=0$. However, 
the second average $\langle r(t)\circ d\widetilde W_2(t)\rangle$ is not zero
as one could naively expect. With the help of the Furutsu--Novikov theorem
\cite{Furutsu,Novikov} we obtain
\begin{eqnarray}
\langle r(t)\circ \widetilde \xi_2(t)\rangle &=\int_{-\infty}^t ds \;
\langle \widetilde \xi_2(t) \widetilde \xi_2(s)\rangle \;\Big \langle \frac {\delta r(t)}
{\delta \widetilde \xi_2(s)}\Big \rangle  \nonumber\\
&= \frac {1}{2}\;\Big\langle\frac {\delta r(t)}{\delta \widetilde
\xi_2(s)}\Big \rangle_{{\big |_{{\scriptstyle{s=t}}}}}=
\frac {1}{2}\frac {\sigma}{\sqrt{2}}\;\langle R(t)\rangle \;.
\end{eqnarray}
Note that the averages here are with respect to realizations of $\widetilde
\xi_2$.

We now observe that the resulting equation for $R$ does not depend on $\widetilde
\xi_2$, therefore $\langle R \rangle=R$, and obtain 
\begin{eqnarray}\label{2RS}
dR=\left[\left(a+\frac{\sigma^2}{4}\right)R-R^3\right]dt
+\frac{\sigma}{\sqrt{2}}
R\circ d \widetilde W_1(t)  \,,
\end{eqnarray}
from which the threshold $a_c(2)=-{\sigma^2}/{4}$ follows.

The system (\ref{2R},\ref{2r}) in Stratonovich sense can be written in a
compact form as $d\boldsymbol{\rho} = \mathbf f(\boldsymbol{\rho})dt +
\sum_{j=1,2}\mathbf g^{(j)}(\boldsymbol{\rho})\,\circ\,d\widetilde W_{j}(t)$,
where  $\boldsymbol{\rho} = (R,r)^T$. 
The equivalent Ito system is 
$d\boldsymbol{\rho} = (\mathbf f + 1/2\sum_{j}
\partial_{\boldsymbol{\rho}}\mathbf g^{(j)}   \mathbf
g^{(j)})dt  + \sum_{j}\mathbf
g^{(j)}\,d\widetilde W_{j}(t)$, where the drift term is
modified by the Ito shift; $\partial_{\boldsymbol{\rho}}\mathbf g^{(j)}$ is
the shorthand of a  Jacobian, cf. e.g. \cite{BBB00}. 
For our system we have $\mathbf g^{(1)}= \sigma / \sqrt{2} \;
\boldsymbol{\rho}$ and  $\mathbf g^{(2)}= \sigma/\sqrt{2}\;(r,R)^T$. The Ito shift
amounts to $\sigma^2/2 \;\boldsymbol{\rho}$ so that the equivalent Ito version of
(\ref{2R}) reads
\begin{eqnarray}
dR=\left[\left(a+\frac{\sigma^2}{2}\right)R\!-\!R^3\!-\!3R r^2\right]\!dt
\!+\!\frac{\sigma}{\sqrt{2}}\!
\left(R \,d \widetilde W_1(t)\!+\!r \,d\widetilde W_2(t)\right).
\end{eqnarray}
Again, $R$ feels only the average of the terms associated with the fast process
$r$, we have $\langle r^2\rangle=0$ but now also the second average  vanishes
since in the Ito calculus $\langle r(t) d\widetilde W_2(t)\rangle=\langle
r(t)\rangle \langle d \widetilde W_2(t)\rangle =0$ which results in
\begin{eqnarray}
dR=\left[\left(a+\frac{\sigma^2}{2}\right)R-R^3\right]dt 
+\frac{\sigma}{\sqrt{2}}
R \,d \widetilde W_1(t).
\end{eqnarray}
This is indeed the Ito eqivalent to Eq. (\ref{2RS}) which can be seen
observing that in  the single variable case the Ito shift is simply $1/2\;g'\;
g = \sigma^2/4\;R$. 

For arbitrary $N$ the same procedure leads to
$a_c(N)=-({\sigma^2}/{2})\left(1-{1}/{N}\right)$ as obtained above.
\end{appendix}

\FloatBarrier

\section*{References}

\bibliographystyle{iopart}

\begin{thebibliography}{99}
\bibitem{SODESim} 
We used the stochastic Runge-Kutta method in a version proposed in 
\cite{KloedenPlaten92, BB96} which is an explicit algorithm  for stochastic
ordinary differential equations  in the Stratonovich sense that converges with
weak order $1$, cf. \cite{BB98}.  The step size was always
$0.01$.
\bibitem{MaxCorr} Trajectories of $x_i(t)$ are numerically determined by an adapted
Heun method \cite{KloedenPlaten92} with time step $0.01$ for different control
parameters $a \gtrapprox a_c$. Values of $a$ for which $x_i(t)$ comes very close to
zero are discarded to exclude that $a<a_c$. After the  transient period the steady
state temporal averages are build.   Supposing a power law $\langle R \rangle
\propto (a-a_c)^{\beta}$ the data are displayed for different test values of $a_c$
in a double logarithmic plot. After a visual control we determine $a_c$ by
maximizing the linear correlation coefficient as proposed in  \cite{GM99}, $\beta$
is the slope of the corresponding line.
\bibitem{ShortTimeRelax} 
The transition points $a_c$ are determined observing numerically the behaviour
of $R(t)$ for an initial state where all $x_i(0)=R_0$ with $R_0^3 \ll R_0$ for
a short period of time such that the distribution of the  $x_i(t)$ becomes not
too broad. Then, the evolution is essentially governed by the linear part of
the Langevin-equation of $R(t)$, cf. \ref{StratIto}.  Trajectories of
$x_i(t)$ are generated by a stochastic Runge-Kutta method with step $0.01$, \cite{SODESim}. 
Generically, for $a<a_c(N)$ the trajectory of $R(t)$ will relax towards zero, whereas for  $a>a_c(N)$ it
will increase provided $R_0$ was smaller than the saturation value. Finally,
the estimates of $a_c(N)$ obtained from different trajectories are averaged. A
similar procedure has been exploited in \cite{BLMKB02} for $N \to \infty$; cf.
also \cite{SSG-O07}.
\bibitem{EulerMaruyama} The Euler-Maruyama method, see
\cite{KloedenPlaten92,Mar55}, is the stochastic equivalent of the explicit
Euler method and has strong order $1/2$ convergence. The step size was $0.01$.
\end{thebibliography}

\vfill
\end{document}